\let\TeXyear\year
\documentclass[doublecol,figures]{epl2}

\let\year\TeXyear
\usepackage[T1]{fontenc}
\usepackage[utf8]{inputenc}

\usepackage{xcolor}
\usepackage{tikz}
\usetikzlibrary{calc,arrows.meta}

\usepackage{graphicx}
\usepackage{amsmath,amssymb}
\usepackage{booktabs}
\usepackage{dblfloatfix}
\usepackage[ruled,vlined]{algorithm2e}
\usepackage{hyperref} 

\definecolor{Blue2}{RGB}{26,145,184}

\definecolor{mypurp}{RGB}{169,83,246}
\definecolor{myyello}{RGB}{235,198,85}
\definecolor{mygree}{RGB}{130,208,67}
\definecolor{myred}{RGB}{202,91,46}

\definecolor{myblue_policy}{RGB}{30,144,255} 
\definecolor{myred_policy}{RGB}{218,59,37} 
\definecolor{myblend}{RGB}{199,131,158}
\definecolor{myblendGR}{RGB}{166,150,57}

\DeclareRobustCommand\upcirc{%
	\begin{tikzpicture}[baseline=-0.5ex]
		\fill[mypurp] (0,0) circle (0.12cm);
		\draw[-{Classical TikZ Rightarrow[length=.8mm]}, thick, black] (0,0) -- (90:0.22cm);
	\end{tikzpicture}%
}
\DeclareRobustCommand\downcirc{%
	\begin{tikzpicture}[baseline=-0.5ex]
		\fill[myyello] (0,0) circle (0.12cm);
		\draw[-{Classical TikZ Rightarrow[length=.8mm]}, thick, black] (0,0) -- (270:0.22cm);
	\end{tikzpicture}%
}
\DeclareRobustCommand\rightcirc{%
	\begin{tikzpicture}[baseline=-0.5ex]
		\fill[mygree] (0,0) circle (0.12cm);
		\draw[-{Classical TikZ Rightarrow[length=.8mm]}, thick, black] (0,0) -- (0:0.22cm);
	\end{tikzpicture}%
}
\DeclareRobustCommand\leftcirc{%
	\begin{tikzpicture}[baseline=-0.5ex]
		\fill[myred] (0,0) circle (0.12cm);
		\draw[-{Classical TikZ Rightarrow[length=.8mm]}, thick, black] (0,0) -- (180:0.22cm);
	\end{tikzpicture}%
}

\newcommand{\hbn}{\hat{\mathbf{n}}}

\title{Harnessing swarms for directed migration of interacting active particles via optimal global control}
\shorttitle{Harnessing swarms for directed migration of active particles} 

\author{Chiara Calascibetta\inst{1} \and La\"etitia Giraldi\inst{1} \and J\'er\'emie Bec\inst{1,2}}
\shortauthor{C.\ Calascibetta \etal}

\institute{                    
  \inst{1} Centre Inria d'Université Côte d'Azur, Calisto - 2004 route des Lucioles, 06560 Valbonne, France\\
  \inst{2} Université Côte d'Azur, CNRS, Institut de Physique de Nice - 17 rue Julien Lauprêtre, 06200 Nice, France
}

\abstract{
This study investigates the use of global control strategies to enhance the directed migration of swarms of interacting self-propelled particles confined in a channel.  Uncontrolled dynamics naturally leads to wall accumulation, clogging, and band formation due to the interplay between self-organization and confinement. This work explores whether a uniform global control, such as magnetic field acting on all particles, can optimize collective transport. Using a discrete Vicsek-like model, it is found that simple global alignment controls, optimized via reinforcement learning, efficiently suppress unfavorable configurations and significantly increase the net particle flux along a prescribed channel direction. These results highlight that coarse, system-level observations are sufficient to achieve near-optimal control, even in regimes with strong fluctuations or partial ordering.
}

\begin{document}

\maketitle

\section{Introduction}
Several practical applications rely on understanding the collective dynamics of swarms of interacting self-propelled particles in confined geometries, such as narrow pipes or channels. Examples include targeted drug delivery, where micro-swimmers must navigate constricted vascular networks~\cite{mitchell2021engineering,wang2012nano,li2017micro}, and micro-robotic systems for environmental monitoring in constrained aquatic environments~\cite{wang2023micro}. The need for greater efficiency, accuracy, and control in these contexts raises questions on how to predict, manipulate, and exploit the collective behavior of micro-swimmers under geometric confinement.

A central challenge in these settings is the design of global control strategies capable of prescribing targeted overall behaviors~\cite{yang2024machine}. In contrast to local control mechanisms, which rely on continuous feedback on individual agents and quickly become impractical for large-scale problems~\cite{calascibetta203CommPhys}, global inputs, such as external fields~\cite{akolpoglu2025navigating} or uniform flow forcing, provide  a realistic means of coordinating many autonomous particles simultaneously. This viewpoint aligns with the broader literature on swarm control, where broadcast or coarse-grained inputs are used to guide large groups without requiring detailed agent-level sensing or communication~\cite{brambilla2013swarm,winfield2018swarm,tovey2025swarmrl}. Such strategies can trigger transitions toward favorable self-organized patterns, synchronize motion across the swarm, and enhance the capacity of micro-swimmers to perform tasks such as coherent navigation, obstacle avoidance, or targeted transport with higher efficiency and reliability. Developing scalable global control paradigms is therefore essential, not only for leveraging the full potential of micro-swimmer collectives but also for advancing theoretical descriptions of active matter~\cite{chate2020dry}. Reinforcement learning and optimal control have been successfully applied in reduced, low-dimensional active systems~\cite{biferale2019zermelo,jayakumar2020machine,daddi2021hydrodynamics,monthiller2022surfing,calascibetta2023EPJE,elkhiyati2023steering}, typically involving few agents or relying on detailed local information. However, the effective state-space dimension of multi-agent systems grows exponentially with the number of particles, rendering such approaches computationally prohibitive and highlighting the need for genuinely scalable global control strategies.

Optimizing the displacement of an ensemble of locally interacting active particles poses several fundamental challenges. How does a global control, such as an external field or uniform forcing, reshape the long-term statistics of the swarm? Can such global interventions reliably select large-scale collective patterns and steer the group efficiently, even in the presence of geometric constraints or self-organized structures? More broadly, to what extent can control remain effective when it relies only on a coarse, system-level information rather than detailed microscopic knowledge of individual agents? To begin addressing these issues, the present work focuses on a minimalistic model for swarm dynamics, designed to isolate the essential mechanisms by which global controls interact with local self-organization. 

\section{Model}
We follow the discrete active-Potts dynamics introduced in~\cite{peruani2011traffic} and consider self-propelled particles evolving on a two-dimensional square lattice with spacing $h$. The domain has a channel geometry: it is periodic with length $L_x$ in the $x$ direction and confined by two walls at $y= \pm L_y/2$. Each lattice site can host at most one particle, capturing steric repulsion, and the total number of particles $N_{\rm p}$ is fixed, yielding a constant density $\rho$. Each particle $k$ carries a discrete orientation $\hbn_k \in \{\upcirc, \downcirc, \leftcirc, \rightcirc \}$. At a swimming rate $\lambda_{\rm S}$, particle $k$ hops to the neighboring site in the direction $\hbn_k$, provided the target site is empty. Confinement is enforced using reflective boundary conditions at $y = \pm L_y/2$: if a particle oriented toward a wall attempts to cross it, its orientation is inverted, i.e, $\hbn_k \mapsto - \hbn_k$.
In addition to translational motion, particles reorient. Particle $k$ adopts a new orientation $\hbn^\prime_k$ with rate $\lambda_{\rm O} \exp \left[g\,\sum_{\ell\sim k} \hbn^\prime_k \cdot \hbn_{\ell}\right]$, where the sum runs over the particles $\ell$ occupying the four nearest-neighbor sites. The alignment strength $g$ controls the tendency of particles to align with their neighbors: for $g = 0$, reorientation is purely random, whereas larger $g$ promotes local alignment.

\begin{figure}[ht]
\includegraphics[width=\columnwidth]{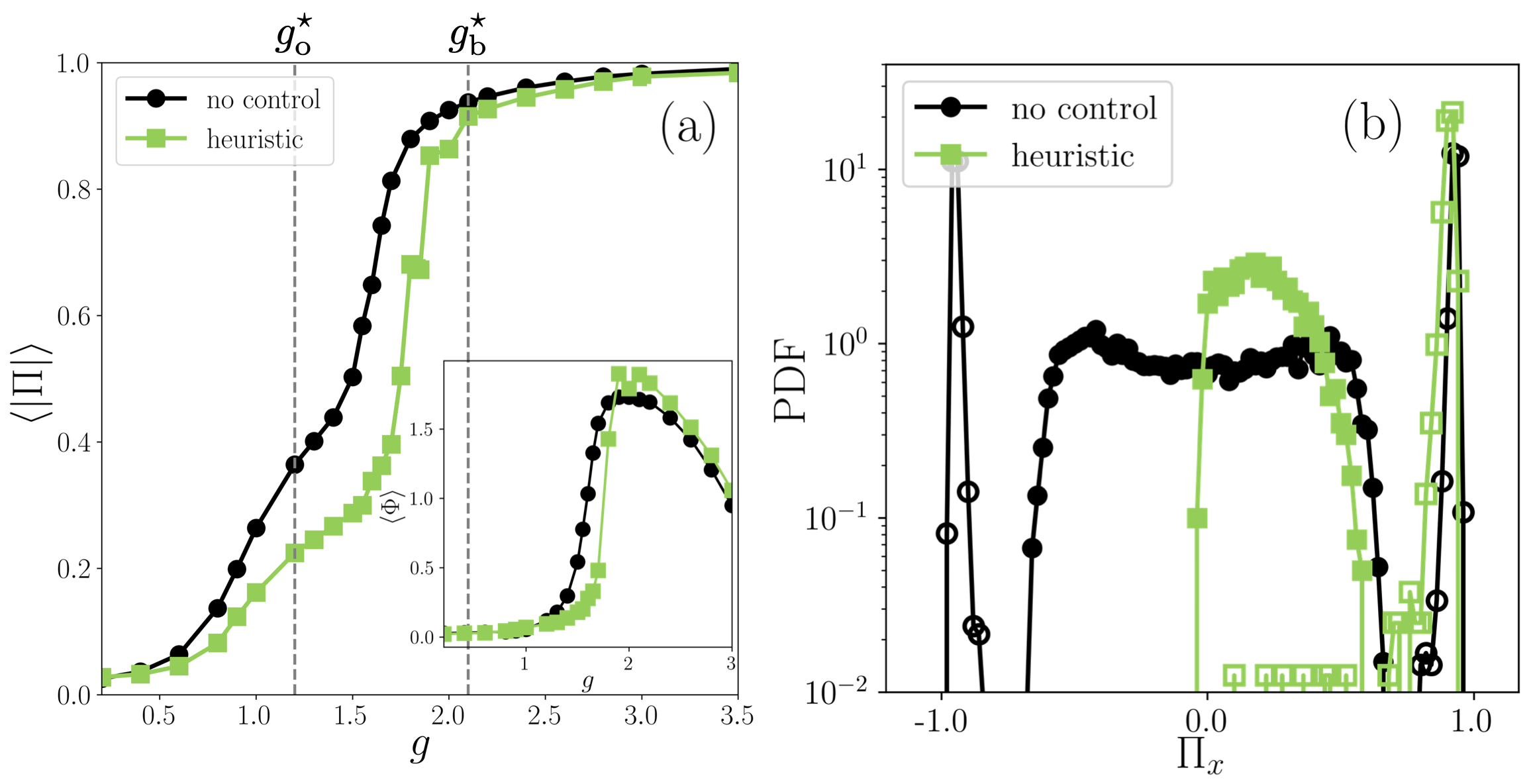}
\vspace{-25pt}
\caption{\textit{Comparison of swarm statistics without control and under the heuristic policy.} 
\textbf{(a)} Global order parameter $\langle|\boldsymbol{\Pi}|\rangle$ as a function of the alignment strength $g$ for the uncontrolled system ($\bullet$) and under the heuristic control (\textcolor{mygree}{\tiny$\blacksquare$}). \textit{Inset:} averaged flux magnitude, $\langle | \Phi|\rangle$, for the uncontrolled system and averaged flux, $\langle \Phi\rangle$, under control.
\textbf{(b)}  Probability density function (PDF) of the mean $x$-component of the orientation $\langle \Pi_x \rangle$, for the uncontrolled system at $g\simeq g_{\rm o}^\star$ ($\circ$) and $g  > g_{\rm b}^\star$ ($\bullet$), and under the heuristic control at $g\simeq g_{\rm o}^\star$ (\textcolor{mygree}{\tiny$\blacksquare$}) and at $g  > g_{\rm b}^\star$ (\textcolor{mygree}{\tiny$\square$}). }
\label{fig:model}
\end{figure}

This lattice model reduces to a continuous-time Markov chain that can be integrated efficiently using Gillespie-type algorithms (see~\cite{peruani2011traffic,calascibetta2024effects}). Despite its minimalism, it captures key hallmarks of active matter, including transitions between different collective patterns as $g$ varies. A key observable is the global order parameter $|\boldsymbol{\Pi}|$, where
\begin{equation}
    \boldsymbol{\Pi} = \frac{1}{N_{\rm p}}  \sum_{k=1}^{N_{\rm p}} \hbn_k\,,
\end{equation}
is the mean orientation of the swarm. It distinguishes three representative regimes (Fig.~\ref{fig:model}a): (i) a disordered phase for $g \lesssim g_{\rm o}^\star$ where only small, short-lived clogs form, (ii) a transitional regime for $g_{\rm o}^\star<g<g_{\rm b}^\star$ characterized by intermittent formation of clogs and bands, and (iii) a fully ordered phase for $g>g_{\rm b}^\star$, where the system exhibits long-lived bands aligned with the channel wall. These bands are metastable and the collective orientation can occasionally flip between $\leftcirc$ and $\rightcirc$. As discussed in~\cite{calascibetta2024effects}, the threshold values $g_{\rm o}^\star$ and $g_{\rm b}^\star$ depend on system parameters. 

Our goal is to design a global control, applied uniformly to all particles, that maximizes net migration in the $x>0$ direction. Achieving this is non-trivial: band formation does not inherently promotes transport, as bands may align against the desired direction or become overly dense, leading to jamming and reduced flux~\cite{calascibetta2024effects}. It is therefore unclear whether a global control can bring about alternative patterns that outperform spontaneous band formation for $g>g_{\rm b}^\star$, or even generate efficient transport at lower $g$, where no ordering naturally emerge.

\

To implement a global control mechanism, we apply an external action at discrete times $t_n = n\,\omega_{\rm c}^{-1}$, where $\omega_{\rm c}$ is the control frequency. At each control time, one of three actions is chosen: align particles horizontally, align them vertically, or leave them unchanged. Specifically:\\
$\bullet$~\textbf{Horizontal action} (${\color{myblue_policy}\boldsymbol{\leftrightarrow}}$): horizontally oriented particles remain unchanged; vertically oriented ones ($\upcirc$, $\downcirc$) become horizontal ($\leftcirc$ or $\rightcirc$, with probability $1/2$).\\
$\bullet$~\textbf{Vertical action} (${\color{myred_policy}\boldsymbol{\updownarrow}}$): vertically oriented particles remain unchanged; horizontally oriented ones ($\leftcirc$, $\rightcirc$) become vertical ($\upcirc$ or $\downcirc$, with probability $1/2$).\\
$\bullet$~\textbf{No action} ($\varnothing$): all particle orientations are unchanged.\\[4pt]
A practical interpretation is that particles are magnetized and subjected at control times to a spatially uniform magnetic field, which tends to align them with the field direction. The probabilistic rule in the lattice model results from coarse-graining continuous orientations into the four discrete directions. For instance, a particle labeled as ``right-oriented'' corresponds to a continuous orientation vector $\hbn = (\cos\theta,\sin\theta)$ with $\theta \in {[} - \pi/4, +\pi/4 {]}$. Under a vertical control, those with $\theta>0$ lie closer to the upward direction and those with $\theta<0$ closer to downward, which is modeled discretely via a $50\%$ probability of turning up or down. A similar interpretation applies when horizontally reorienting  vertical particles.  

The control frequency should be chosen with care. If  $\omega_{\rm c}$ is too large, repeated control pulses may override intrinsic self-organization, preventing collective patterns from developing. Conversely, if  $\omega_{\rm c}$ is too low, the imposed action may be washed out by natural fluctuations, allowing the system to drift away from the targeted configuration before the next intervention. Ideally, $\omega_{\rm c}^{-1}$ should lie between two characteristic timescales: the relaxation time required for the swarm to respond to a control pulse and approach a desired collective pattern, and the typical lifetime of such a pattern before it destabilizes due to fluctuations. These timescales are not known a priori and depend sensitively on the physical parameters of the system, most notably on the alignment strength $g$, which strongly affects pattern formation and stability. In practice, we fix the control frequency to $\omega_{\rm c}= h \lambda_{\rm S} / L_x$, whose inverse corresponds to the time needed for an isolated particle to swim along the channel, thereby to be possibly captured by a collective pattern. This choice provides a physically meaningful baseline that does not privilege any particular self-organized structure. A more detailed analysis of the dependence on $\omega_{\rm c}$ is provided later.

The control problem consists in determining a policy that assigns an action to each state of the system. An effective policy should rely on the most informative yet compact description of the swarm, without requiring full knowledge of the whole state space. For instance, the global order parameter $|\boldsymbol{\Pi}|$ alone is insufficient for maximizing the net displacement in the $x>0$ direction: it quantifies the degree of alignment but does not distinguish whether particles are oriented left or right. To capture this directional information, we instead consider the $x$-component $\Pi_x = \boldsymbol{\Pi}\cdot\hat{\boldsymbol{x}}$ of the mean orientation vector $\boldsymbol{\Pi}$. The sign of $\Pi_x$ directly indicates whether the swarm is predominantly oriented toward the left or the right, and thus whether a given configuration hinders or favors the desired migration. This information alone leads to a simple \textit{heuristic policy}:
\begin{equation} \label{eq:heuristic}
\pi_{\mathrm{heur}} = \begin{cases} {\color{myred_policy}\boldsymbol{\updownarrow}}, & \Pi_x < 0,\\ {\color{myblue_policy}\boldsymbol{\leftrightarrow}}, & \Pi_x > 0. \end{cases}
\tag{Heur}
\end{equation}
In words, when $\Pi_x < 0$, particles are chiefly aligned opposite to the target direction and a vertical action is applied to disrupt such states. Conversely, when $\Pi_x > 0$, indicating alignment with the desired flux, a horizontal action is applied to reinforce this favorable orientation. This deterministic policy therefore alternates between suppressing unfavorable states and strengthening favorable ones. 

To quantify the performance of a given control policy, we define the net displacement per unit time as
\begin{equation}\label{eq:flux}
    \varphi_{n,m} := \frac{h}{N_\mathrm{p}}\frac{\!{\left[N_{\rm right}(t_n,t_{m} ) - N_{\rm left}(t_n,t_{m})\right]}}{t_{m} - t_n}\,,
\end{equation}
where $N_{\rm right}(t_n,t_{m})$ and $N_{\rm left}(t_n,t_{m})$ denote, respectively, the total number of rightward and leftward particle hops occurring in the time interval $(t_n,t_{m})$. This quantity measures the net transport velocity per particle induced by the control. For convenience, we further introduce a dimensionless flux, $\Phi = \varphi_{n,n+10}/(h\lambda_{\rm O})$,  sampled every 10 control times $\omega_{\rm c}^{-1}$. For any fixed policy, its performance is then assessed by averaging $\Phi$ over long times, after discarding an initial transient.

\begin{table}[h!]
\centering
\caption{Dimensionless system parameters.}
\label{tab:parameters}
\begin{tabular}{p{3cm}cp{1.8cm}}
\hline
Resolution  & $N_y = L_y/h$ & $11$ \\ \hline
Aspect ratio & $N_x/N_y$ & $50/11\approx 4.5$ \\ \hline
Alignment strength & $g$ & $0$ --- $3.5$\\ \hline
Péclet number & $\lambda_\mathrm{S}/\lambda_{\rm O}$ & $100$\\ \hline
Density & $\rho = N_{\rm p}/(N_x\,N_y)$ &  $0.3$\\ \hline
Control frequency & $\omega_{\rm c}/\lambda_{\rm O}$ & 2 \\ \hline
\end{tabular}
\end{table}
Table~\ref{tab:parameters} provides the parameters used in the simulations. Lengths are scaled by the lattice spacing $h$, and times by the reorientation rate $\lambda_{\rm O}$. The resolution $N_y$ and aspect ratio $N_x /N_y$ are chosen to provide a domain large enough for several collective structures while remaining computationally affordable. The density $\rho$ is set to avoid both the dilute regime, where interactions are negligible, and the jammed limit, where steric effects dominate. The Péclet number fixes migration and alignment to act on comparable timescales for $g=\mathcal{O}(1)$. The alignment strength $g$ is varied to probe disorder, transitional, and band-forming regimes. Finally, the influence of the control time $\omega_{\rm c}^{-1}$, initially set by the time an isolated particle needs to cross the channel, is examined later.

Figure~\ref{fig:model}a compares the phase transition of the global order parameter $|\boldsymbol{\Pi}|$ as a function of $g$ for both the uncontrolled system and the system driven by the heuristic policy. The main effect of the control seems to shift of the transition to larger values of $g$, thereby reducing the range over which ordered behavior occurs. However, as observed in the inset of Fig.~\ref{fig:model}a, this is associated to an improved flux in the $x>0$ direction in the controlled case. To facilitate interpretation, the mean flux amplitude $\langle|\Phi|\rangle$ is shown for the uncontrolled system (as $\Phi$ averages to zero due to symmetry between left- and right-oriented states~\cite{calascibetta2024effects}), while the exact mean flux $\langle\Phi\rangle$ is reported for the controlled system. The heuristic policy notably increases the mean flux at large $g$, where macroscopic order can be exploited to enhance net transport. Figure~\ref{fig:model}b shows the probability density function (PDF) of the streamwise polarization $\Pi_x$ for two representative values of the alignment strength: $g = 1.2 \simeq g_{\rm o}^\star$, near the onset of order, and $g = 2.1 > g>g_{\rm b}^\star$, deep in the ordered regime. In both cases, the heuristic strategy induces a preferential positive alignment along the channel, producing sightly narrower tails than in the uncontrolled case.

Phenomenologically, the heuristic policy works by skewing the distribution of $\Pi_x$ toward the desired direction. At low $g$, where the system is close to disorder, applying the horizontal action whenever $\Pi_x>0$ reinforces rightward fluctuations, creating a net positive bias. At high $g$, where spontaneous band formation can produce leftward-oriented clusters, the vertical action disrupts these unfavorable configurations, effectively destroying left-aligned bands and allowing right-aligned structures to dominate. In this way, the policy both generates a directional bias in disordered regimes and selectively removes counterproductive patterns in highly ordered regimes.

\begin{figure*}[t!]
\includegraphics[width=\textwidth]{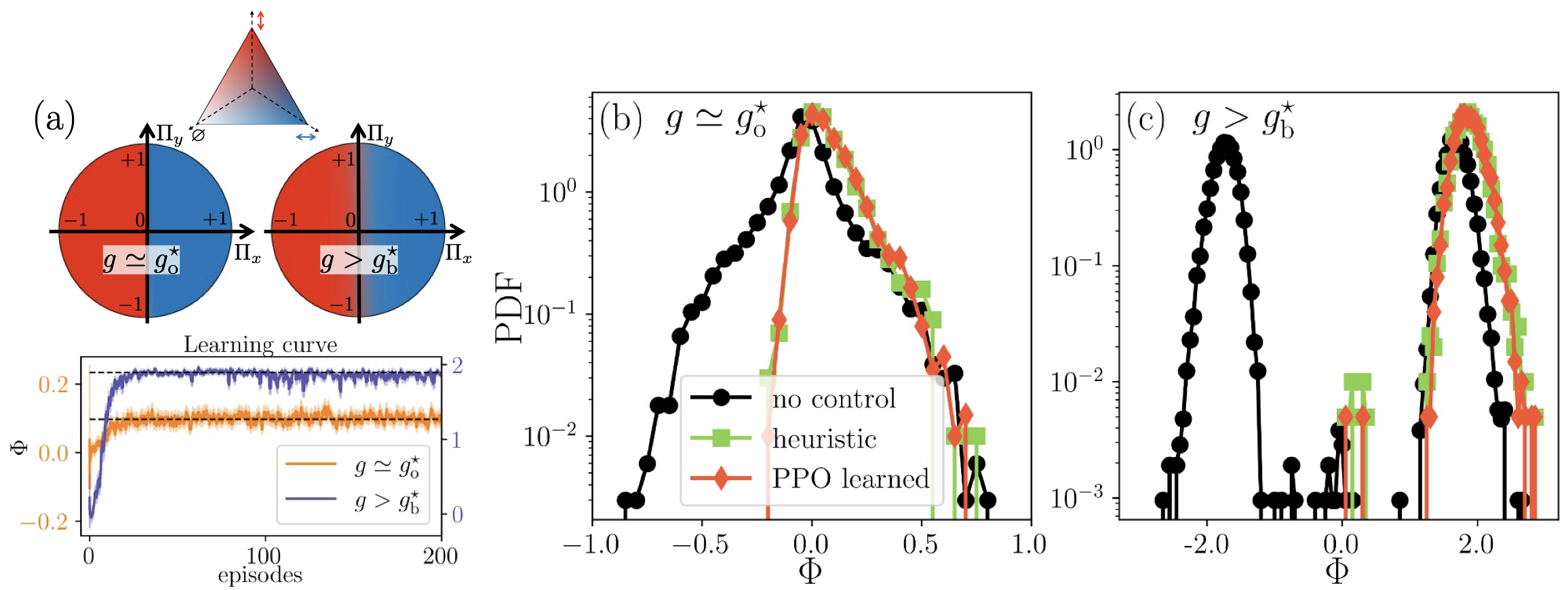}
\vspace{-25pt}
\caption{\textit{Disordered ($g\simeq g_{\rm o}^\star$) and fully ordered ($g >g_{\rm b}^\star$) regimes.}
\textbf{(a)} \textit{Top:} PPO policy map in the $(\Pi_x,\Pi_y)$ plane for $g = 1.2$ and $g = 2.1$. Colors indicate action probabilities: red = vertical (\textcolor{myred_policy}{$\boldsymbol{\updownarrow}$}), blue = horizontal (\textcolor{myblue_policy}{$\boldsymbol{\leftrightarrow}$}), white = no action ($\varnothing$), with intermediate shades denoting stochastic mixtures. \textit{Bottom:} learning curves, showing the streamwise fluxes concatenated across training episodes. Black dashed lines mark the performance of the heuristic policy. 
\textbf{(b-c)} Probability density functions of the streamwise flux $\Phi$  at $g\simeq g_{\rm o}^\star$ (b) and $g>g_{\rm b}^\star$ (c), comparing the uncontrolled system ($\bullet$), the heuristic policy (\textcolor{mygree}{\tiny$\blacksquare$}), and PPO (\textcolor{myred}{$\blacklozenge$}).}
\label{fig:extremes}
\end{figure*}
The control framework introduced above can be naturally formulated in terms of reinforcement learning (RL), which provides a framework to optimize  global control. At each control step $n$, the system is described by an observed state $s_n \in \mathcal{S}$. A global action $a_n\in\{\textcolor{myred_policy}{\boldsymbol{\updownarrow}}, \textcolor{myblue_policy}{\boldsymbol{\leftrightarrow}}, \color{black}\varnothing\}$ is then applied according to a policy $\pi(a|s)$, which defines the probability of selecting action $a$ given the state $s$. For instance, the heuristic policy introduced above corresponds to $s = \mathrm{sign}(\Pi_x) \in \{-1,1\}$ and to the deterministic choice $\pi(a|s) = \delta(a -\textcolor{myred_policy}{\boldsymbol{\updownarrow}})\,\delta(s+1) + \delta(a - \textcolor{myblue_policy}{\boldsymbol{\leftrightarrow}})\,\delta(s-1)$. The immediate reward associated with each step is defined as the normalized flux between two controls
\begin{equation}
    r_n := {\varphi_{n, n+ 1}}/({ h\lambda_{\rm S}})\,.
\end{equation}
This normalization makes the reward dimensionless and of order one.  
The overall goal of control is to maximize the cumulative reward, i.e., to identify policies that drive the swarm preferentially toward positive $x$. This objective is formalized through the value function
\begin{equation}\label{eq:def_value}
V_\pi(s) = \sum_{n\ge0} \gamma^n\,\mathbb{E}\left[r_n\middle| s_0 = s\right]\,,
\end{equation}
which quantifies the expected discounted cumulative reward starting from state $s$ under policy $\pi$. In principle, determining the optimal policy requires maximizing $V_\pi(s)$ for all $s \in \mathcal{S}$. In practice, however, the high dimensionality of the system render direct optimization intractable, motivating the use of data-driven RL approaches.

In the following, we implement RL optimization and compare its performance with the heuristic policy. Key questions include: Is the heuristic strategy near-optimal? Can such a coarse, global state description is sufficient to significantly influence swarm dynamics? More broadly, can an optimized global control enhance performance and consistently enforce a positive flux across disordered, transitional, and ordered regimes?

\section{Results}
We now assess the performance of reinforcement learning within the control framework introduced above. Global control policies are optimized using the proximal policy optimization (PPO) algorithm, a state-of-the-art on-policy actor-critic method. PPO combines policy-gradient updates with stabilization techniques such as clipped policy loss and generalized advantage estimation (GAE)~\cite{mecanna2025critical}. In our implementation, we use as state variables the continuous values of the two components of the mean orientation vector $\boldsymbol{\Pi} = (\Pi_x,\Pi_y) = (1/N_{\rm p}) \sum_{k=1}^{N_{\rm p}} \hbn_k$, and the action set $\{\color{myred_policy}\boldsymbol{\updownarrow},\color{myblue_policy}\boldsymbol{\leftrightarrow},\color{black}\varnothing\}$. The optimization target is the cumulative streamwise flux defined in Eq.~\eqref{eq:def_value}. Unlike the heuristic policy, which relies on a deterministic threshold of $\Pi_x$, PPO can exploit continuous state information and exploit correlations in the dynamics that are not captured by simple heuristics. We also explored extended state representation, including the global bond-orientational order parameter $\Psi_4$, which encodes the mean number of nearest neighbors and clustering~\cite{calascibetta2024effects}, and $y$-resolved observables accounting for wall proximity. Such additions could, in principle, compensate for the coarseness of global averages. However we observed no qualitative improvement and in all regimes, PPO converged to policies with performance comparable to those with the $(\Pi_x,\Pi_y)$ state description.

\begin{figure*}[t!]
\includegraphics[width=\textwidth]{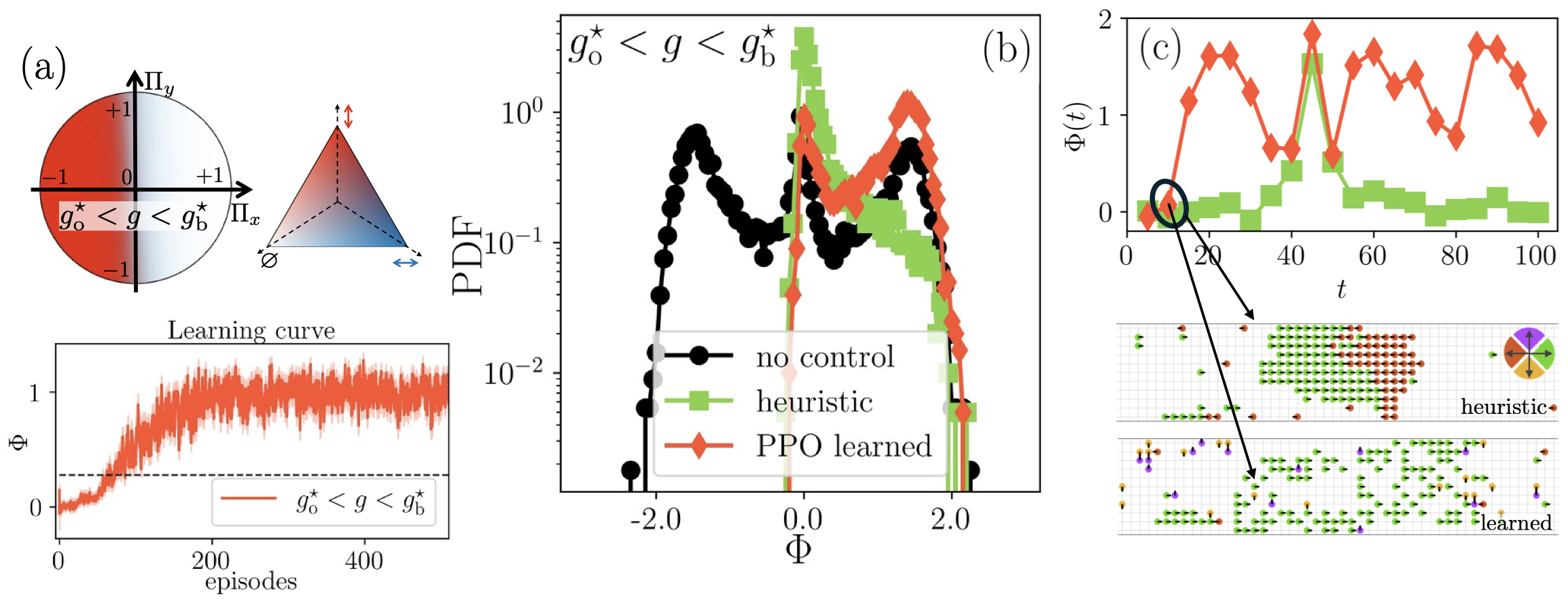}
\vspace{-25pt}
\caption{\textit{Transitional regime ($g_{\rm o}^\star<g<g_{\rm b}^\star$).} \textbf{(a)} \textit{Top:} PPO policy map in the $(\Pi_x,\Pi_y)$ plane for $g = 1.6$, using the same color code as Fig.~\ref{fig:extremes}a. \textit{Bottom:} associated learning curve, with the heuristic performance shown as a dashed line. \textbf{(b)} PDFs of the flux $\Phi$, comparing the uncontrolled system ($\bullet$), the heuristic policy (\textcolor{mygree}{\tiny$\blacksquare$}), and the policy learned from PPO (\textcolor{myred}{$\blacklozenge$}). \textbf{(c)} \textit{Top:}  time series of $\Phi$ for heuristic (\textcolor{mygree}{\tiny$\blacksquare$}) and PPO (\textcolor{myred}{$\blacklozenge$}) starting from the same initial condition. \textit{Bottom:} representative snapshots at $t\simeq 15$, showing particle orientations.}
\label{fig:transient}
\end{figure*}
Because the control task has no natural terminal state, the problem is formulated as an infinite-horizon optimization. Simulations are partitioned into episodes of fixed duration $T_{\rm{max}} = 2000\,\omega_{\rm c}^{-1}$,  each consisting of 2000 control steps with a fixed policy. At the end of each episode, the policy and value networks are updated via gradient ascent, using multiple epochs of shuffled mini-batch to reduce temporal correlations and improve sample efficiency. This procedure is repeated until convergence. A complete list of hyperparameters is provided in Tab.~\ref{tab:ppo_main}.

We first assess the performances of PPO in the two extreme regimes of the system: near the onset of collective order ($g\simeq g_{\rm o}^\star$) and deep in the ordered phase ($g> g_{\rm b}^\star$). Figure~\ref{fig:extremes}a shows the resulting policy maps accross the ($\Pi_x, \Pi_y$) plane. In both cases, PPO converges toward an almost deterministic threshold rule: the action is essentially a function of the sign of $\Pi_x$, with probabilities saturating to either the vertical or the horizontal action. As already stated, we also tested richer state representations by augmenting $(\Pi_x,\Pi_y)$ with additional order parameters, but this produced no appreciable differences: the learned policy remained effectively identical and achieved the same performance. The learning curves (Fig.~\ref{fig:extremes}a, bottom) show fast convergence, typically after $\sim 50$ training episodes. 

\begin{table}[!b]
\centering
\caption{PPO setup (condensed).}
\label{tab:ppo_main}
\begin{tabular}{@{}ll@{}}
\toprule
\text{Algorithm} & PPO (clipped surrogate)  \\
\text{Discount / GAE}        & $\gamma=0.8$ \;/\; $\lambda=0.75$ \\
\text{Clip param.}           & $\epsilon=0.2$ \\
\text{Actor network}             & 2$\times$40 (ELU), softmax output \\
\text{Critic network}            & 2$\times$100 (ELU), linear output \\
\text{Optimizers}            & Adam (actor/critic, separate) \\
\text{Rollout length}              & $2000$ \\
\text{Minibatches}              & $4$ \\
\text{Epochs}              & $4$ \\
\text{Episodes}              & $>400$ \\
\text{Learning rate actor}        & $3\times10^{-5}$ --- $3\times10^{-4}$ \\
\text{Learning rate critic}        &  $10^{-4}$ --- $10^{-3}$ \\
\bottomrule
\end{tabular}
\end{table}
Once a policy is obtained, it is frozen and used to generate statistically stationary trajectories for performance evaluation. Figures~\ref{fig:extremes}b and c show the PDFs of the streamwise flux $\Phi$ for $g\simeq g_{\rm o}^\star$ and $g> g_{\rm b}^\star$. In both cases, PPO and the heuristic shift the distribution toward positive flux, relative to the uncontrolled system, and their performances are indistinguishable within sampling fluctuations. This indicates that the heuristic policy is already optimal at these two extreme regimes. Still, the underlying mechanism differs in each regime. Near $g_{\rm o}^\star$, horizontal alignment is fragile and rarely sustained; the heuristic's vertical-horizontal switching simply introduces a mild statistical that PPO cannot significantly improve. For $g> g_{\rm b}^\star$, the swarm naturally forms robust bands, and the heuristic effectively suppresses the negative-orientation bands while reinforcing the positive ones, thereby enhancing the mean flux relative to the uncontrolled case. Two conclusions follow. First, PPO algorithm consistently converges to a nearly deterministic strategy, assigning almost all probability to a single action in the states it visits. Second, very coarse, global information (essentially the sign of $\Pi_x$) already suffices to achieve optimal performance in both the disordered and fully ordered phase.

Convergence to the heuristic policy is no longer guaranteed in the transitional regime $g_{\rm o}^\star<g<g_{\rm b}^\star$, where partial order coexists with strong fluctuations~\cite{calascibetta2024effects}. Figure~\ref{fig:transient}a shows the corresponding PPO policy map and the learning curve. In this regime, PPO converges more slowly than at the extremes, yet consistently surpasses the heuristic once trained. The resulting policy is again nearly deterministic and takes the simple \textit{disruption-only} form:
\begin{equation}
    \pi_{\mathrm{disr}} = \begin{cases} {\color{myred_policy}\boldsymbol{\updownarrow}}, & \Pi_x < 0,\\ \varnothing, & \Pi_x > 0, \end{cases}
    \tag{Disr}
\end{equation}
that is, the vertical action is applied only to disrupt left-oriented configurations. We also tested again richer state inputs, but these did not alter the resulting learned policy nor improve performance.

Figure~\ref{fig:transient}b reports the PDFs of the flux $\Phi$ at a representative value $g = 1.6$. The heuristic controller still enforces a net positive flux, but mainly by eliminating negative events; its distribution remains narrowly peaked near zero. The uncontrolled system, by contrast, exhibits a secondary positive peak at higher flux that the heuristic fails to stabilize. PPO departs from the heuristic and achieves clear gains: it suppresses the negative tail while recovering, and slightly amplifying, the high-flux positive peak of the uncontrolled case, leading to a larger mean flux. This can be interpreted as follows: the disruption-only policy prevents the system from locking into compact right-oriented bands, which tend to jam due to volume exclusion. Instead, it allows the swarm to self-organize into more mobile, loosely packed right-oriented structures. Snapshots from simulations started from the same initial condition (Fig.~\ref{fig:transient}c) illustrate this contrast: the heuristic produces denser clusters, whereas the disruption-only policy preserves more open configurations that translate more effectively. This explains why the policy learned from PPO outperforms the heuristic in the transitional regime.

\begin{figure}[!b]
    \centering
    \includegraphics[width=1\linewidth]{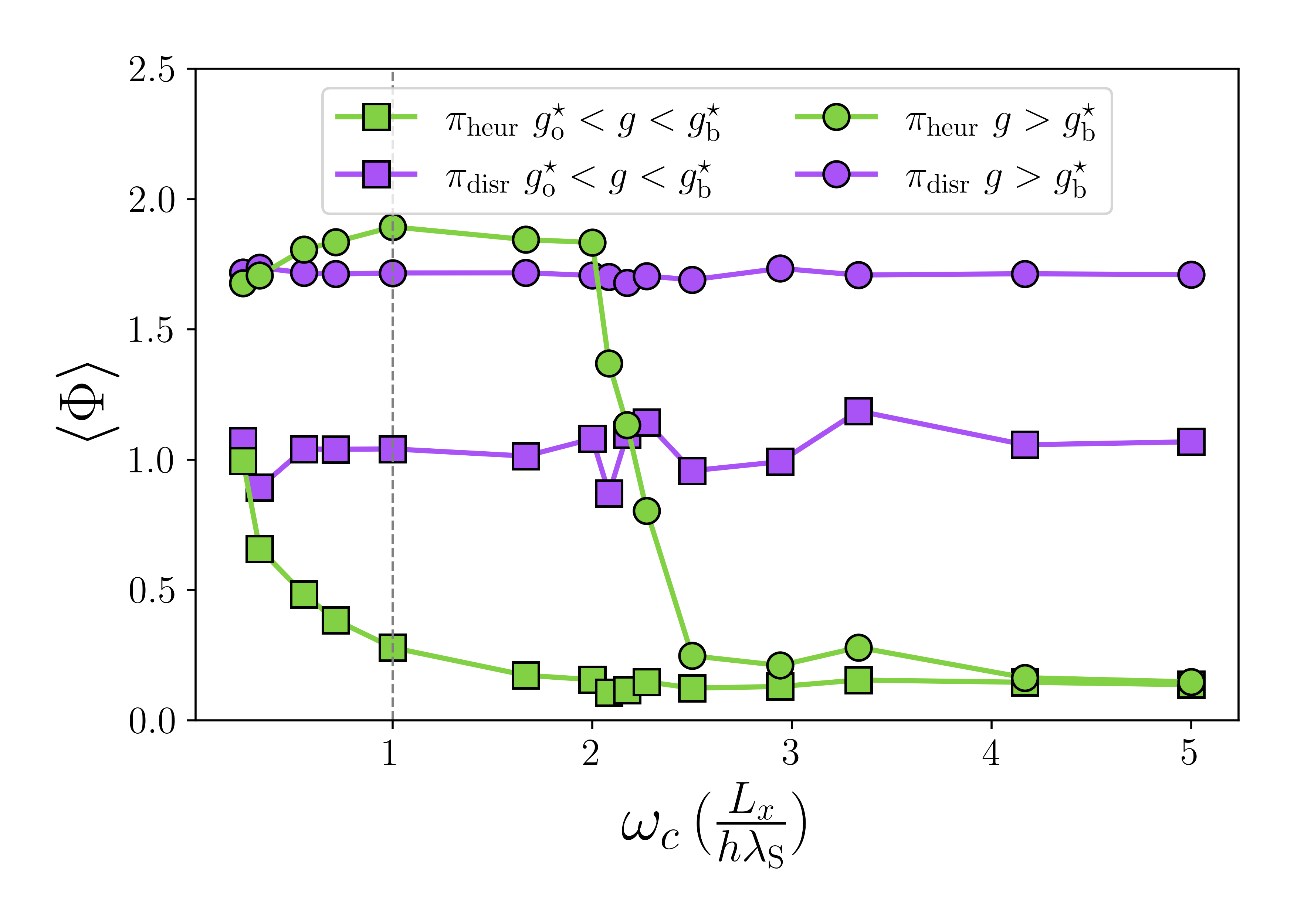}
    \vspace{-25pt}
    \caption{\textit{Dependence of mean streamwise flux on the control frequency $\omega_{\rm c}$}. Results are shown for the heuristic policy $\pi_{\rm heur}$ in the the transitional regime $g_{\rm o}^\star< g <g_{\rm b}^\star$ (\textcolor{mygree}{\tiny$\blacksquare$}) and the ordered regime $g > g_{\rm b}^\star$ (\textcolor{mygree}{$\bullet$}), and for the disruption-only policy $\pi_{\rm disr}$ (\textcolor{mypurp}{\tiny$\blacksquare$} and \textcolor{mypurp}{$\bullet$}). The vertical dashed line marks the reference control frequency used throughout the manuscript, $\omega_{\rm c} = h\lambda_{\rm S}/L_x$.}
\label{fig:dependence_tauc}
\end{figure}
Applying one of the two optimal deterministic policies, $\pi_\mathrm{heur}$ or $\pi_\mathrm{dirs}$, outside its favorable regime while varying the control frequency $\omega_{\rm c}$ clarifies the trade-offs underlying effective control (Fig.~\ref{fig:dependence_tauc}). We explored a broad range of frequencies, $0.25\,h\lambda_{\rm S} /L_x \le \omega_{\rm c}\le 5\,h\lambda_{\rm S} /L_x$. For small $g$ in the disordered regime, performance increases monotonically as $\omega_{\rm c}$ increases: both policies become essentially equivalent because actions occur before any collective structure can form (not shown). At intermediate $g$ in the transitional regime, the disruption-only policy $\pi_\mathrm{dirs}$ systematically outperforms the heuristic for all tested $\omega_{\rm c}$ (squares in Fig.~\ref{fig:dependence_tauc}). In this regime, allowing the system to evolve freely when $\Pi_x>0$ is always beneficial and additional horizontal actions tend to degrade the flux. The two policies are comparable only at very small $\omega_{\rm c}$, where correlations between successive controls are effectively suppressed. In contrast, for large $g$ in the ordered regime, the situation reverses. When the control frequency is too large ($\omega_{\rm c} \gtrsim 2h\lambda_{\rm S}/L_x$), the heuristic policy is penalized because frequent interventions disrupt the translation of the right-oriented band (circles in Fig.~\ref{fig:dependence_tauc}). For $\omega_{\rm c} \gtrsim 2h\lambda_{\rm S}/L_x$, however, the heuristic consistently outperforms the disruption-only policy, down to very small frequencies where long-time correlations again vanish. This confirms that exploiting band formation requires leaving enough time for right-oriented bands to form and persist, and avoiding overly frequent actions that would fragment them.

\section{Discussion and outlook}
Our results point to a simple overall picture: the best global control always balances two effects, making use of the swarm's natural tendency to self-organize, and enforcing streamwise alignment when needed. The right balance between these two mechanisms depends sensitively on the alignment strength $g$. In the extreme regimes, either strongly disordered or strongly ordered, the vertical-horizontal heuristic policy $\pi_{\rm heur}$, which forces horizontal reorientation when the swarm already has a favourable mean orientation and vertical reorientation when it does not, proves the most effective. When $g$ is small, orientational fluctuations behave essentially like noise, so imposing horizontal alignment provides the only reliable directional bias. When $g$ is large, the dynamics itself already favours long-lived, right-oriented bands; horizontal actions then strengthen these structures, leading to enhanced transport. By contrast, in the transitional regime at intermediate values of $g$, order is intermittent and fluctuating~\cite{calascibetta2024effects}. The most efficient policy switches to the disruption-only strategy $\pi_{\rm disr}$. By acting only when the mean swarm orientation is unfavourable and remaining passive otherwise, this policy allows transient right-oriented structures to develop without letting them become overly compact. Remarkably, this simple rule consistently outperforms the heuristic across control frequencies, even when the controller is given access to richer descriptors  such as clustering measures or wall-distance information.

Taken together, our results illustrate how a single global control can reshape both the emerging patterns (band compactness/orientation) and the long-time statistics of an active swarm, without needing to access individual particles. They also show that the learned policies are almost entirely deterministic; the small amount of randomness that remains after learning can be attributed to imperfect convergence of the PPO algorithm. This near-deterministic character suggests that the underlying control landscape might be rather simple, perhaps even close to convex, though establishing this would require a more formal analysis of $\pi \mapsto V_\pi$ in Eq.~(\ref{eq:def_value}). In fact, the nature of the optimal solutions is reminiscent of classical bang-bang controls, which typically arise in such simplified landscapes~\cite{kirk2004optimal}.

Another striking conclusion is that very coarse global observables suffice. Using only the sign of the mean streamwise orientation $\mathrm{sign}\,(\Pi_x)$, performs just as well as policies that rely on two-component polarization or additional geometric features. This can be rationalized using a simple mutual-information argument. The mutual information between the state space $\mathcal{S}$ and the action set $\mathcal{A}$, which quantifies how much knowing the system state reduces uncertainty about the action, is $I(\mathcal{S};\mathcal{A}) = H(\mathcal{A}) - H(\mathcal{A}|\mathcal{S})$, where $H$ denotes the Shannon entropy~\cite{cover2005elements}. This immediately leads to the upper bound  $I(\mathcal{S};\mathcal{A}) \le H(\mathcal{A}) \le \log_2(3) \simeq 1.58$ bits, since the action space contains only three choices and its entropy is maximized for an equiprobable distribution. Thus, the information that is transmitted from state to action is intrinsically limited by the action alphabet, and no matter how rich the state representation becomes, the effective channel capacity cannot increase unless the action set is expanded.

Broadly, this work contributes to the ongoing effort to understand how simple, global inputs can steer the collective behaviour of active systems confined in narrow geometries. The key idea is to replace detailed, agent-level actuation with weak fields acting simultaneously on all particles, and to substitute fine-grained microscopic sensing with global, aggregate measurements. This viewpoint is appealing from a physics perspective, as it highlights the interplay between self-organization and external driving, and from a practical perspective, as it reflects realistic constraints in microfluidic or microrobotic applications, where actuation is limited and sensing is inherently coarse.

Several natural next steps follow from this study. A first step is to move beyond the lattice model and explore continuous systems that include steric forces and hydrodynamic interactions, such as Vicsek-type models or squirmers, in order to understand how global control couples to ambient flows and to swimmer-flow feedback. It is also worth broadening the notion of ``global action''. Even weak spatial modulations of a uniform field, or fields with tunable amplitude or periodicity, could reveal whether slight spatial or temporal structure leads to measurable gains. On the sensing side, it becomes necessary to replace idealised observables with noisy signals and to test robustness across densities, geometries and parameter uncertainty, including whether learned strategies transfer effectively to new conditions. Another key question concerns timing: learning not just \textit{what} to do but \textit{when} to intervene may clarify the relation between the decision rate and the swarm's intrinsic self-organization timescales. 

Overall, the long-term goal is to turn the principles of global control into practical strategies for directing swarms in realistic microfluidic and robotic environments, where interventions must be sparse, sensing capabilities are limited, and success depends on working with, rather than against, the swarm's own collective dynamics.

\acknowledgments
We thank Zakarya El Khiyati and Michele Buzzicotti for useful discussions.
Computational resources were provided by the OPAL infrastructure from Universit\'{e} C\^{o}te d'Azur.
This work was supported by the Agence Nationale de la Recherche (Grants No. ANR-21-CE45-0013 \& ANR-21-CE30-0040-01).

\textit{Data availability statement:} The data that support the
findings of this study are available from the authors upon reasonable request.

\end{document}